\documentclass[11pt]{article}
\usepackage[a4paper,margin=1in]{geometry}
\usepackage{graphicx}
\usepackage{hyperref}
\usepackage{amsmath}
\usepackage{listings}
\usepackage{color}
\usepackage{caption}

\title{\textbf{MHD\_Macroscopic\_ball\_analysis: An Open Toolkit for Analyzing Magnetic Particle Rotation in Viscous Media}}
\author{Roberts Treize\\
Laboratory of Magnetic Soft Materials, University of Latvia\\
\texttt{roberts.treize@edi.lv}}

\date{May 2025}

\begin{document}
\maketitle

\begin{abstract}
This paper presents \texttt{MHD\_Macroscopic\_ball\_analysis}, an open-source Python toolkit developed to study the rotational dynamics of magnetized spherical particles immersed in viscous media under externally applied rotating magnetic fields. The toolkit provides extraction of the magnetic moment orientation from grayscale video frames using image processing, and enables detailed comparisons with synchronized magnetic field vector data. We outline the motivation, core algorithms, implementation, and validation of the toolkit in the context of macroscopic experiments simulating magnetohydrodynamic (MHD) particle behavior. The full source code and documentation are freely available on \href{https://github.com/streboreziert/MHD_Macroscopic_ball_analysis}{GitHub}, with a citable version archived on \href{https://doi.org/10.5281/zenodo.15530413}{Zenodo}.
\end{abstract}

\section{Introduction}
Rotational dynamics of magnetic particles play a central role in magnetic hyperthermia, microfluidics, and soft robotics\cite{cebers2003dynamics,cebers2005chaotic,Peiravi2022, Szwed2024,Pearce2013}. Experimental observation of such systems at the microscale is often constrained by resolution and control limitations. To address this, our laboratory developed a macroscopic model using a magnetized ball in glycerin under a controlled rotating field.

To support data processing from this setup, we created \texttt{MHD\_Macroscopic\_ball\_analysis}: a Python-based toolkit to extract and analyze the 3D orientation of a spherical magnet from image sequences and compare it to the applied field. This toolchain has supported both theoretical validation and dynamic visualization.

\section{System Overview}
The experimental setup includes a PLA sphere containing a magnet, suspended in glycerin and actuated by a three-axis Helmholtz coil system. The ball's rotation is filmed using a monochrome camera, while the applied magnetic field is recorded in parallel.

The toolkit consists of two main components:
\begin{itemize}
    \item \texttt{process\_ball\_images.py}: Detects the sphere in each frame, identifies its dividing line, and computes the orientation vector.
    \item \texttt{align\_coil\_vector\_with\_ball.py}: Aligns extracted vectors with time-synchronized magnetic field data and computes dynamic angles such as lag and inclination.
\end{itemize}

\section{Implementation}

\subsection{Image-Based Vector Extraction}
Using OpenCV \cite{bradski2000opencv}, the script detects circular contours and traces the hemisphere dividing line\cite{canny1986computational,qin2020u2net,otsu1979threshold,duda1972use,gonzalez2002digital}. From three points on the sphere, a normal vector is constructed to estimate the direction of the internal magnetic moment. The time series of these vectors is saved in a structured format.

\subsection{Coil Field Alignment}
The second script compares the extracted moment vectors to the applied magnetic field values. By computing the angle between these vectors over time, it determines lag angle, inclination, and azimuth. Visualization and CSV output are generated for further analysis.

\section{File Formats}

\subsection*{Input Images}
\begin{itemize}
    \item Format: Grayscale PNG or JPG
    \item Resolution: 2448 × 2048 or any
    \item Naming: Sequential (e.g., \texttt{frame\_0000.png}, \texttt{frame\_0001.png})
\end{itemize}

\subsection*{Field Data}
CSV with columns: \texttt{timestamp}, \texttt{Bx}, \texttt{By}, \texttt{Bz} (units in Tesla or microtesla)

\subsection*{Output}
\begin{itemize}
    \item \textbf{Moment vectors}: \texttt{timestamp, mx, my, mz}
    \item \textbf{Analysis results}: \texttt{lag\_angle, inclination, azimuth}
\end{itemize}

\section*{Availability}
All code and documentation are available at \href{https://github.com/streboreziert/MHD_Macroscopic_ball_analysis}{GitHub}, with a citable archive at \href{https://doi.org/10.5281/zenodo.15530413}{Zenodo}.

\section*{Citation}
If you use this toolkit in your research, please cite:
\begin{quote}
Roberts Treize. \textit{MHD\_Macroscopic\_ball\_analysis (v1.0.0)}. Zenodo. 2025.\\
\url{https://doi.org/10.5281/zenodo.15530413}
\end{quote}

\section*{Acknowledgments}
This code was developed as part of the magnetic particle dynamics project at the University of Latvia under the under the guidance of Dr. phys. Jānis Cīmurs, Associate Professor at the University of Latvia.

\section*{License}
This project is released under the MIT License.

\bibliographystyle{plain}  
\bibliography{main}  

\end{document}